\documentclass[11pt]{article}

\usepackage[ruled,vlined]{algorithm2e}

\usepackage{amsmath,geometry}
\usepackage{fullpage}
\usepackage{natbib}
\usepackage{lscape}
\usepackage{setspace}
\usepackage{authblk}
\usepackage{graphics,graphicx}
\usepackage{appendix}
\usepackage{subfigure}
\usepackage{url}
\usepackage{floatrow}
\usepackage{amsmath,amssymb,amsthm}
\usepackage{bm}
\usepackage{graphicx}
\usepackage{booktabs}
\usepackage{comment}

\DeclareMathOperator*{\argmin}{argmin}

\date{}

\title{Self-Validated Ensemble Models for Design of Experiments}

\author[1]{\small Trent Lemkus}
\author[2]{\small Christopher Gotwalt}
\author[1]{\small Philip Ramsey}
\author[3]{\small Maria L. Weese}
\affil[1]{\small Department of Mathematics \& Statistics, University of New Hampshire, Durham, NH}
\affil[2]{\small SAS, Cary, NC 27513 }
\affil[3]{\small Department of Information Systems \& Analytics, Miami University, Oxford, OH}

\begin{document}


\maketitle
\doublespace

%
%

 


\begin{abstract}
\textcolor{black}{One of the possible objectives when designing experiments is to build or formulate a model for predicting future observations. When the primary objective is prediction, some typical approaches in the planning phase are to use well-established small-sample experimental designs in the design phase (e.g., Definitive Screening Designs) and to construct predictive models using widely used model selection algorithms such as LASSO. These design and analytic strategies, however, do not guarantee high prediction performance, partly due to the small sample sizes that prevent partitioning the data into training and validation sets, a strategy that is commonly used in machine learning models to improve out-of-sample prediction. In this work, we propose a novel framework for building high-performance predictive models from experimental data that capitalizes on the advantage of having both training and validation sets. However, instead of partitioning the data into two mutually exclusive subsets, we propose a weighting scheme based on the fractional random weight bootstrap that emulates data partitioning by assigning anti-correlated training and validation weights to each observation. The proposed methodology, called Self-Validated Ensemble Modeling (SVEM), proceeds in the spirit of bagging so that it iterates through bootstraps of anti-correlated weights and fitted models, with the final SVEM model being the average of the bootstrapped models. We investigate the performance of the SVEM algorithm with several model-building approaches such as stepwise regression, Lasso, and the Dantzig selector. Finally, through simulation and case studies, we show that SVEM
generally generates models with better prediction performance in comparison to one-shot model selection approaches.}

\end{abstract}
%
%

Weighted Bootstrap, Lasso

\section{Introduction}\label{sec:intro}

\textcolor{black}{In industrial and scientific experiments, predictive models are often used in process or product optimization \cite{kincl2005application, attala2021application}, calibration \cite{bondi2012effect}, and sensitivity analysis \citep{mahanthesh2021flow}. Experiments for building predictive models are well-grounded in experimental design theory, with central composite designs or CCD's \citep{CCD}, Box-Behnken designs or BBD's
\citep{BBD}, and definitive screening designs or DSD's \citep{DSD} being the gold standard for small-sample experiments with multiple factors. Model building typically proceeds using well-established and widely available\footnote{Available in popular statistical software such as JMP, SAS, and R} model selection algorithms, such as stepwise regression, Lasso \citep{Lasso}, and the Dantzig selector \citep{candes2007dantzig}.}

\textcolor{black}{Despite the availability of design and analytic recommendations for building predictive models using experimental data, generated models have been documented to encounter several issues in practice. First, it has been shown that without going through systematic model validation, some model selection algorithms, such as Lasso and the Dantzig Selector, produce poor to mediocre predictive models \citep{Mein1, tibshirani2014praise}. In building machine learning (ML) models, the typical approach for improving a model's predictive capability is by performing out-of-sample validation, which proceeds by partitioning the data set into training and validation sets. However, experimental data are typically characterized by small run sizes due to resource constraints; thus, partitioning experimental data into a subset for training and another for validation may be at odds with the objective of improving predictive accuracy through validation \citep{breiman_instability, Yuan_Var_Selection}. This is because holding out one or more runs could substantially change the estimated relationship between the predictors and the response, render some model terms inestimable, or result in worse predictive performance \citep{CV_DoE}.}

\textcolor{black}{Secondly, in designed experiments, assumptions about the model form (e.g., addition of higher-order polynomial or interaction terms) are often a concern in practice, where experimenters are often uncertain if more complex relationships exist among factors. Traditional design of experiments also subscribes to the principle of effect hierarchy i.e., higher-order model terms are more likely to be inactive \citep{box1978statistics}, specifically when the experimental objective is screening or determining which factors greatly impact the response. In the prediction realm, however, it has been shown that smaller, reduced models often yielded higher prediction errors than models with greater complexity \citep{Smucker_RSM}. This is because studies with a prediction objective are geared towards emulating complex functions to predict new observations accurately \citep{Shmueli_Explain_Predict}. Thus, supersaturated models, where the number of model parameters exceed the number of observations, frequently have superior prediction performance on validation data \citep{Belkin_BiasVar_Tradeoff}, a finding that we will corroborate in a later section. Fitting supersaturated models, however, is not standard practice in the analysis of experimental data but is often observed in machine learning.}

\textcolor{black}{From the previous discussion, it is evident that building predictive models using small-sample experimental data could benefit from standard methodologies used in machine learning but due to its small-sample attributes, capitalizing on these ML methodologies, until now, has been implausible. In this research, we propose a novel model building framework for experimental data that facilitates built-in validation of candidate models. The proposed methodology, called Self Validated Ensemble Modeling or SVEM, incorporates self-validation in the model building process without holding out any observations (i.e., the full data set is used at every iteration in the training phase).}

\textcolor{black}{SVEM is motivated by bagging \citep{breiman_bagging}, in that the final model is an ensemble average of a set of bootstrapped models. \cite{breiman_bagging} showed that models that use a single pass through a selection algorithm have, in some cases, been found to suffer from unstable parameter estimates \textcolor{black}{and variation in the selected models.} Since building prediction models from small-sample experimental data is ultimately a model selection problem, instability in parameter estimates could result in ambiguity in determining which model terms should be retained. Consequently, this could result in models with poor predictive performance \citep{breiman_bagging}.}

\textcolor{black}{Bagging, as an alternative to one-pass model selection, has been shown to reduce the impact of model instability and improve prediction performance. Thus, SVEM was developed in its spirit. Similar to bagging, SVEM uses a bootstrapping procedure to construct, for each bootstrap replicate, training and validation sets for fitting and tuning a model. The key difference between the two methods is in the manner that these two sets are constructed. In the bagging procedure, the training set is a resample with replacement from rows of the data set and the validation set consists of rows not selected in the resample. In lieu of partitioning the data into mutually exclusive subsets for training and validation, SVEM uses every observation as a member of both sets through a strategic weighted resampling scheme that resembles the fractionally weighted bootstrap in \cite{Xu_Apps_FRW}.}

\textcolor{black}{SVEM provides a unified framework for addressing the key issues previously described when building predictive models using experimental data from BBD's and DSD's, two commonly used small-sample three-level designs used for building prediction models under the Quality by Design philosophy \citep{Weese_DSD_Screen}. We will show, through simulation, that SVEM outperforms one-shot model selection procedures (e.g., Lasso) with respect to prediction accuracy. Further, we will show that SVEM produces stable parameter estimates.  We also show the effectiveness of SVEM in a real-world manufacturing problem where the objective is to maximize the yield of a biomaterial.}


\textcolor{black}{The rest of the paper proceeds as follows. Section \ref{sec:SVEM} provides details on the theoretical groundwork and practical implementation of SVEM in JMP software \citep{JMP}. Next, we demonstrate the effectiveness of SVEM in improving prediction performance when applied to data from designed experiments using simulation of various design conditions (Section \ref{sec:sim}) and a real-world case study (Section \ref{sec:casestudy}). Finally, conclusions and areas for further research are presented in Section \ref{sec:conclusion}}.

\section{SVEM}\label{sec:SVEM}

\textcolor{black}{In this work we consider predictions from DSDs and BBDs.  These are three-level designs with levels (-1, 0, 1). BBDs are typically fit with a full quadratic (FQ) model, represented below in Equation \eqref{eqn:intro} where $K$ is the number of factors in the experiment. DSDs lack the degrees of freedom to fit a full quadratic model and thus a model selection algorithm (e.g. forward selection) is applied to search for the important terms from Equation \eqref{eqn:intro}.}

\begin{equation}
    \mathbf{Y} = \beta_0 + \sum_i^K \beta_i x_i + \sum_i^K \beta_{ii} x_{ii}^2 + \sum_{i<j}^K \beta_{ij} x_{ij} + \epsilon 
    \label{eqn:intro}
\end{equation}

\textcolor{black}{We define the design matrix as $\mathbf{D_{N,K}}$ where $N$ is the number of observations, and the model matrix by $\mathbf{X_{N,P+1}}$ where $P$ is the total number of columns in the model matrix apart from the intercept. Then we can write Equation \eqref{eqn:intro} using matrix notation as $\mathbf{Y_N} = \mathbf{X_{N,P+1}}\boldsymbol{\beta_{P+1}} + \boldsymbol{\epsilon_N}$. The response or responses to be predicted are represented by a column vector(s) $\mathbf{Y_N}$. Here, $\boldsymbol{\beta}_{P+1}$ is the vector of unknown model parameters and $\boldsymbol{\epsilon_N}$ is the usual vector of random errors. In designs with limited run size, $N$, it is possible that true model is saturated ($P \approx N$) or super-saturated ($P>N$). The limited degrees of freedom in these scenarios lead to poor predictive models especially in cases where the sparsity principle \citep{goos2011optimal} does not hold true.} 


\subsection{Foundations of SVEM}

SVEM blends concepts from bootstrapping with those of ensemble modeling to fit validated predictive models to data from designed experiments. \textcolor{black}{Although we present SVEM in the context of design of experiments}, it is a general algorithm that can be used to fit many \textcolor{black}{different} types of data (e.g.observational data) and predictive modeling strategies. \textcolor{black}{In this work,} our focus is the fitting of linear models of the form shown in Equation \ref{eqn:intro}.

\textcolor{black}{ 
Bootstrapping generates a set of samples of size $N$ by sampling with replacement from an original data set of size $N$. Some of the original observations occur more than once in the re-sampled set, while others do not occur \citep{Efron_Tib_Bootstrap}. Each observation in a bootstrapped re-sample is assigned an integer weight (including 0) based upon how many times that observation was selected for that re-sample. The bootstrap uses sampling with replacement. Therefore, some observations may appear more than once in the re-sample. }

\textcolor{black}{In the SVEM algorithm we use a different bootstrapping procedure with a modified weighting scheme called fractionally weighted bootstrapping (FWB) \citep{Xu_Apps_FRW}. FWB does not sample with replacement. Instead every observation is included in every re-sample. However the fractional-weights assigned to each observation are different in every re-sample. Because FWB does not exclude observations in re-samples it is ideal for data for an experimental design where preserving the full structure of the design in the modeling process is imperative. SVEM uses FWB to assign observations to either a training or a ``self-validation'' sample.} 

\textcolor{black}{Before we define ``self-valdiation''}, consider the usual approach to creating training and validation sets. For fitting models, observations in the training set are assigned a weight of 1, while observations in the validation set are assigned a weight of 0. For validation, these weights are reversed. The integer weights assure that the validation set provides an independent assessment of prediction performance.

\textcolor{black}{In self-validation, used in SVEM, the original data set is the training partition and a copy of the original data set is the ``self-validation'' partition. Each observation in the training set has a twin in the self-validation set. We assign fractional weights in pairs such that if any observation in the training partition is assigned a large weight, the twin in the self-validation partition is assigned a small weight and vice versa. This inverse weighting scheme drives anti-correlation between the two partitions, allowing the self-validation set to function as a \emph{de facto} validation set. Table \ref{Tab:Table_Frac_Weights} shows a typical self-validation data table set-up for a three factor DSD for a single iteration of SVEM. Each time a fractionally weighted bootstrap sample is created, observations are assigned different weights, sometimes high in the training set and sometimes high in the validation set. We use these differently weighted re-samples in the same way a random forest model uses standard bootstrap samples in bagging.}

%

\begin{table}
\begin{tabular}{lrrrrc}
\hline
\textbf{Validation} & \textbf{X1} & \textbf{X2} & \textbf{X3} & \textbf{Y} & \textbf{Fractional Wts} \\ \hline
Training            & 0           & 1           & 1           & 4.03       & 0.388                   \\
Training            & 0           & -1          & -1          & -1.48      & 3.433                   \\
Training            & 1           & 0           & -1          & -0.38      & 0.073                   \\
Training            & -1          & 0           & 1           & -1.82      & 1.732                   \\
Training            & 1           & -1          & 0           & 1.57       & 2.900                   \\
Training            & -1          & 1           & 0           & 1.89       & 1.046                   \\
Training            & 0           & 0           & 0           & 1.21       & 0.067                   \\ \hline
Self-Validation     & 0           & 1           & 1           & 4.03       & 1.135                   \\
Self-Validation     & 0           & -1          & -1          & -1.48      & 0.033                   \\
Self-Validation     & 1           & 0           & -1          & -0.38      & 2.652                   \\
Self-Validation     & -1          & 0           & 1           & -1.82      & 0.195                   \\
Self-Validation     & 1           & -1          & 0           & 1.57       & 0.057                   \\
Self-Validation     & -1          & 1           & 0           & 1.89       & 0.433                   \\
Self-Validation     & 0           & 0           & 0           & 1.21       & 2.734                   \\ \hline
\end{tabular}
\caption{Sample self-validation for a three factor DSD}\label{Tab:Table_Frac_Weights}
\end{table}

\subsection{SVEM Implementation}

\textcolor{black}{In SVEM the fractional weights are random draws from an exponential distribution with mean 1.} The weighting scheme first generates a set of $N$ random uniform (0, 1) weights and then use the inverse probability transform with the exponential distribution is used to generate the fractional weights. Equation \ref{eq_generate_wts} illustrates the computation of the weights, where $U[0,1]$ represents a uniform distribution on the interval (0, 1), and $F$ is the cumulative distribution function for an exponential distribution with mean 1.

\begin{equation} \label{eq_generate_wts}
\begin{split}
Generate:  & u_i \sim U[0,1] \ for \ i=1...N \\
Training \ FWs: \  & w_{T,i} = F^{-1}(u_i) = log(u_i) \ for \ i=1...N \\
Self\mbox{-}Validation \ FWs: \  & w_{V,i} = F^{-1}(1 - u_i) = log(1 - u_i) \ for \  i=1...N
\end{split}
\end{equation}

\begin{figure}[H]
     \centering
     \includegraphics[width=4in]{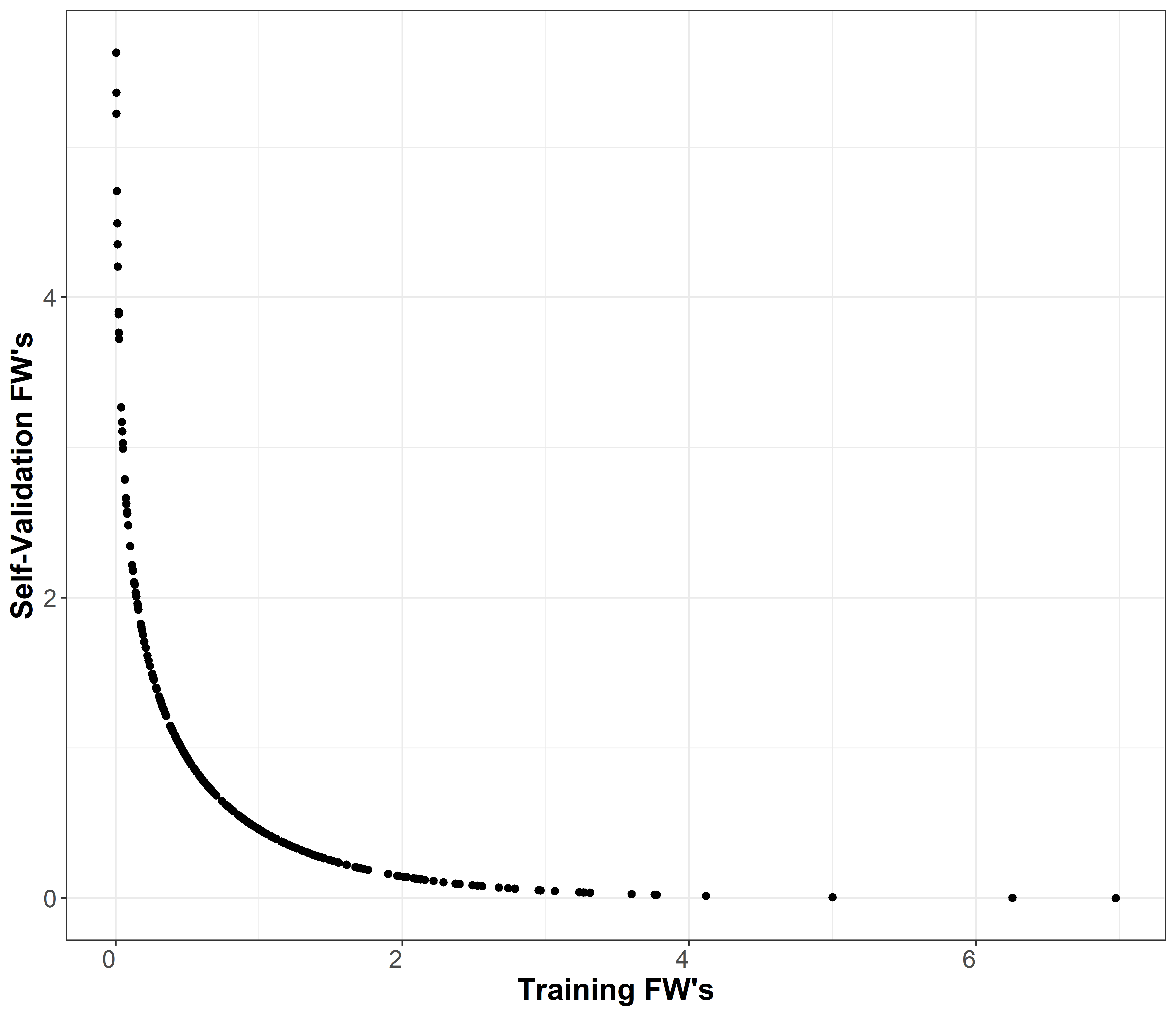}
     \caption{Scatterplot of 250 self-validation FWs and training FWs}\label{fig:FWBbiplot}
\end{figure}

The two sets of weights are assigned in pairs to the training and self-validation partitions (\textcolor{black}{see the last column} in Table \ref{Tab:Table_Frac_Weights}). This relationship is such that any given $w_{T,i}$ assigned randomly to a case in the training data will have a corresponding $w_{V,i}$ assigned to the same case in the \text{self-validation} data. Because of the way the weights are constructed, large values for any given $w_{T,i}$ will have small values for its corresponding $w_{V,i}$ and vice versa. This relationship is represented in Figure \ref{fig:FWBbiplot}, which shows the typical inverse relationship between the training and \textcolor{black}{self-validation} weights that supports the \textcolor{black}{self-validation} approach.

\begin{figure}[H]
     \centering
     \includegraphics[width=\textwidth]{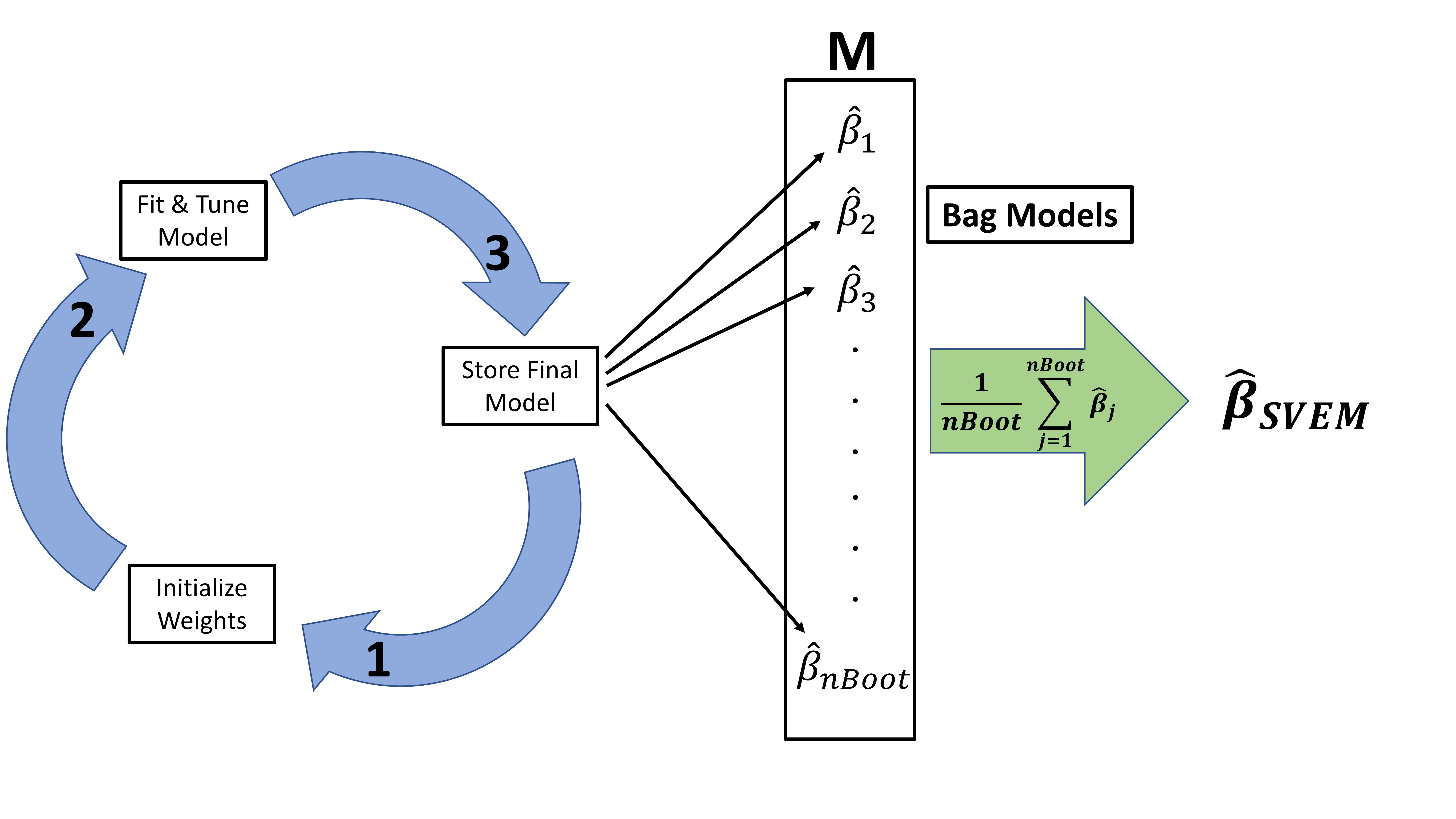}
     \caption{SVEM Workflow Diagram}\label{fig:S_VEM_Diag}
\end{figure}

Once the weights are assigned to the training and self-validation partitions, a model selection algorithm is applied to the training data. 
\textcolor{black}{The model selection algorithm then selects the model with the minimum mean squared prediction error (MSPE) on the self-validation set. The process is repeated for a number of iterations, $nBoot$, specified by the user. At each iteration the chosen model is stored and the final model is computed by averaging the coefficients from each of the $nBoot$ models. For $nBoot$ iterations, a matrix $\mathbf{M}_{nBoot,P+1}$ containing all of the coefficient estimates is created to estimate the final ensemble model. If a specific predictor is not selected on an $nBoot$ iteration, then the associated coefficient value in $\mathbf{M}_{nBoot,P+1}$ is assigned a 0. Algorithm~\ref{algorithim} displays the steps of the SVEM where $\mathbf{M}_i$ represents the $i^{th}$ row in matrix $\mathbf{M}$ and Figure~\ref{fig:S_VEM_Diag} shows a schematic of SVEM modeling process.}

\vspace{0.2cm}
\begin{algorithm}[htb]
\SetAlgoLined
\KwResult{$\hat{\beta}_{SVEM}$}
 \For{$i=1:nBoot$}{
  Generate $\boldsymbol{\tilde{w}_T}, \ \boldsymbol{\tilde{w}_V}$\;
  Fit model $\boldsymbol{f(X,\tilde{w}_{T}|Y)} = \boldsymbol{\hat{f}(.)}$\;
  Calculate $\boldsymbol{SSE^V(\beta) = \argmin\limits_{\beta} \sum_i w_{V,i}(y_i - \hat{f}(X,\beta))^2}$\;
 Select $\boldsymbol{\hat{\beta}} = {\argmin\limits_{\beta} }[SSE^V(\boldsymbol{\beta})]$\;
  $\boldsymbol{\mathbf{M}_i} \gets \boldsymbol{\hat{\beta}}$\;
  }
  Bag$(\boldsymbol{\mathbf{M}}) \longrightarrow  \hat{\beta}_{SVEM}$
 \caption{SVEM}\label{algorithim}
\end{algorithm}   

\section{Simulation Study}\label{sec:sim}        
         
\textcolor{black}{To study the performance of SVEM} we compare the predictive accuracy of SVEM against other model selection algorithms used for prediction. 
We focus on two designs commonly used for prediction in the process industries: definitive screening designs \citep{DSD} (DSDs) and Box-Behnkenn designs \citep{BBD} (BBDs). 
We used DSDs with K = 4 and 8 factors and BBDs with K = 3 and 5 factors. 
In the next section, we outline our simulation protocols,which were all run in JMP Pro version 16 \citep{JMP}.

\subsection{Simulation Protocol}\label{sec:sim_protocol}

We generate a true model $\mathbf{Y} = \mathbf{X}\boldsymbol{\beta}$ \textcolor{black}{according to Equation \eqref{eqn:intro}}, where $\boldsymbol{\beta}$ is randomly sampled from a standard double exponential distribution for the truly active effects in $\mathbf{X}$ the FQ model matrix. The inactive effects are assigned a coefficient of 0. We did not impose heredity to choose the active two-factor interactions and quadratic effects in the true model \textcolor{black}{since not all methods (e.g. Dantzig Selector) can impose heredity in the model selection process}. We chose the number of active effects using three sparsity levels: no sparsity, medium sparsity, and high sparsity. Table \ref{tab:sparsity_sim} gives the percentage of active effects in each scenario. Sparsity refers to the principle that only a subset of predictors, $p\prime$, are relatively small compared to $N$ \citep{Hastie_sparsity_principle}. In a case such as the DSD $K = 8$ with $A = 100\%$, there are 45 active effects (including the intercept) with $N = 21$. This is a case where the true model is very complex and the principle of sparsity is ignored \textcolor{black}{and the model matrix $\mathbf{X}$ is supersaturated}.  

\begin{table}[htb]
\begin{tabular}{ccc}
  & \multicolumn{2}{c}{Percentage of Active Effects} \\\hline
                & DSD                       & BBD                       \\\hline
No Sparsity     & 100\%      & 100\%       \\
Medium Sparsity & 50\%        & 50\%        \\
High Sparsity   & 25\%        & 25\%        \\\hline     
\end{tabular} 

\caption{The percentage of truly active effects for simulation scenarios}
\label{tab:sparsity_sim}
\end{table}


\textcolor{black}{For each design and sparsity combination, we generate 1,000 true response vectors, $\mathbf{Y_i}$ for $i = 1,2,3...,1000$ using $\mathbf{Y}=\mathbf{X}\beta+\mathbf{\epsilon}$ where $\boldsymbol{\epsilon}$ is Gaussian noise. The noise is assigned at two levels: $\mu = 0$ and $\sigma = 1$, $\mu = 0$ and $\sigma = 3$.  For each of the 1,000 true response vectors we apply SVEM and competing model selection methods to select a model and evaluate the predictive accuracy. }

The model selection algorithms we used are: forward selection (FWD), pruned forward selection (PFWD) (a variant of mixed step regression \citep{JMP}), Lasso, \textcolor{black}{the Dantzig Selector} (DS), and SVEM. We implemented SVEM with forward selection (SVEM FWD), pruned forward selection (SVEM PFWD), and the Lasso (SVEM Lasso). For SVEM we used an $nBoot = 200$ to balance performance and computational overload \textcolor{black}{(see Section \ref{sec:sim_results})}. We also included the fit definitive screening method \citep{fitdef}, which is only applicable to DSDs with a FQ. Fit definitive screening has four heredity options: full strong heredity, quadratic heredity only (heredity imposed only on quadratic terms), interactive heredity only (heredity imposed only on interaction terms), and no heredity. We selected the two variants that performed the best overall: full strong heredity (FDS Full) and quadratic heredity only (FDS Quad). 

A common criterion to assess prediction performance is the root mean square prediction error (RMSPE) calculated on the validation set. RMSPE has the benefit of a linear error scale, which is easy to interpret i.e. an error of 10 is twice as bad as an error of 5. We use the RMSPE as our comparison metric for the different models selected by the methods mentioned above. 

\begin{figure}[H]
     \centering
     \begin{subfigure}
         \centering
         \includegraphics[width=0.4\linewidth]{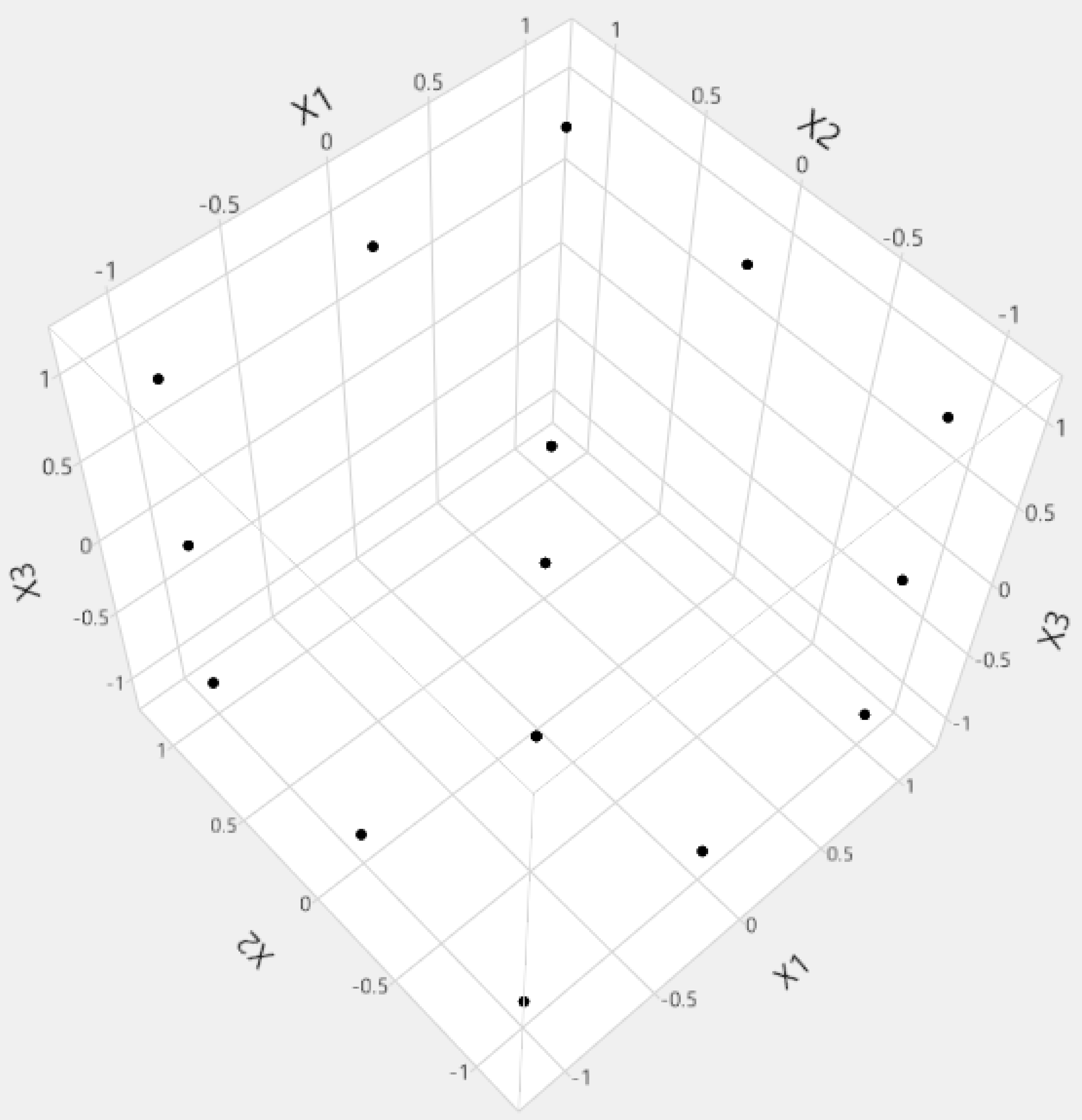}
         \label{fig:DSD_4ME_Region}
     \end{subfigure}
     \begin{subfigure}
         \centering
         \includegraphics[width=0.442\linewidth]{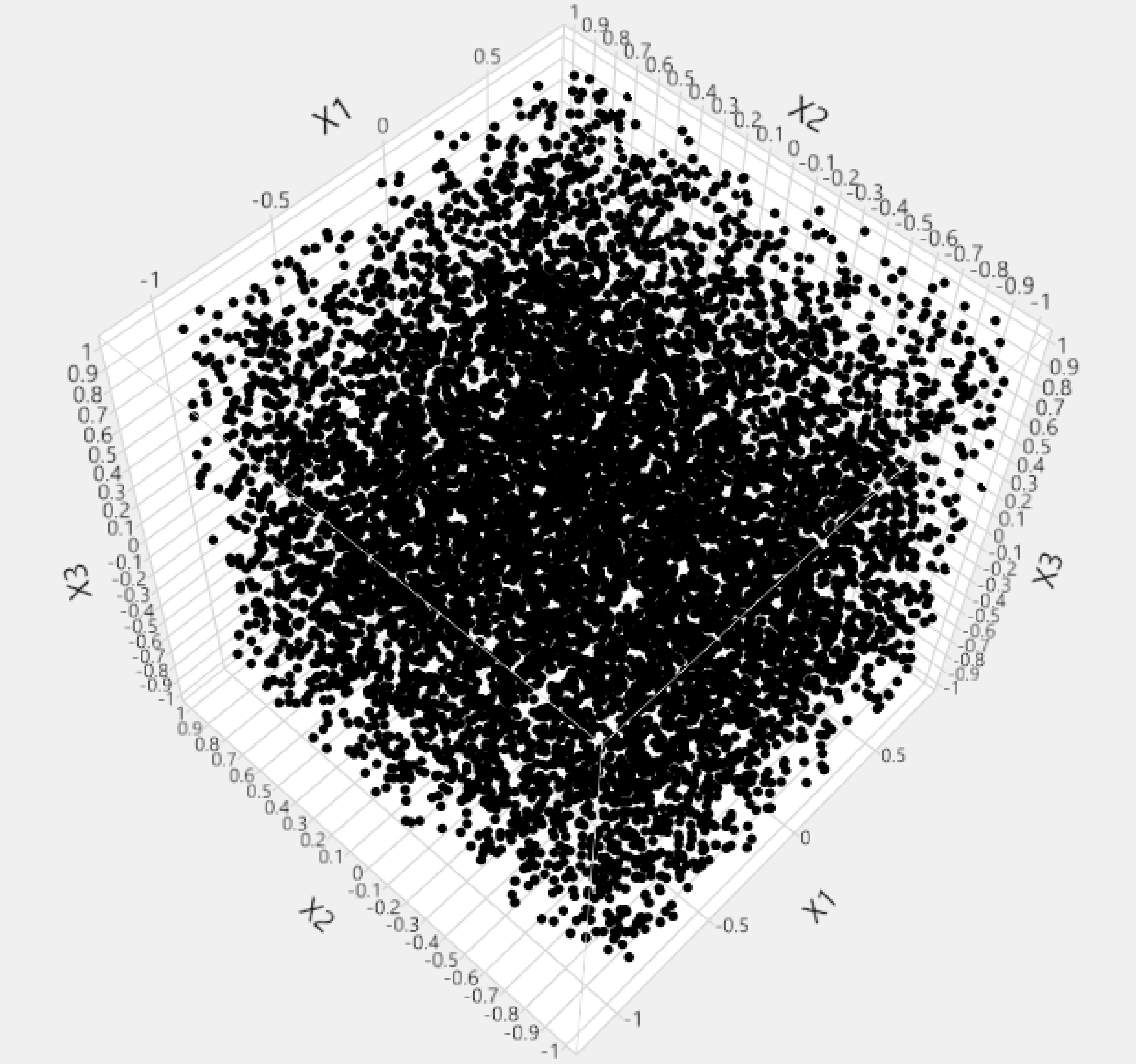}
         \label{fig:SFD_4ME_Surface}
     \end{subfigure}
\caption{Case values in the design region for a DSD with $K = 4$ (left) with its corresponding 1,000 randomly sampled space filling design (right).}
\label{fig:DSD_SFD_Cube}
\end{figure}

\textcolor{black}{In order to ensure we compare predicted values throughout the entire design region and not just on the small sample of experimental design points, we make use of a space filling design \textcolor{black}{(SFD) to generate a response surface. We generate a ``true response" from the ``true model`` and a ``predicted response`` from the selected model $\boldsymbol{\hat{\beta}}$. The SFD is created with 1,000 runs of randomly selected values from the design region \citep{Rosh_SFD} where the design region is scaled to a hypercube of $[-1,1]^K$. To do this we create a matrix $\mathbf{X_{SFD}}$} by randomly sampling each of the $K$ main effects to give elements $x_{i,j}$ a uniform distribution with support $[-1,1]$ where $i = 1,2,...,1,000$ and $j=1,2,...,K$.  We expand  $\mathbf{X_{SFD}}$ to a FQ model matrix $\mathbf{X_{SFD}'}$. We calculate a new ``true response" from the true model by $\mathbf{\tilde{Y}} = \mathbf{X_{SFD}'}\boldsymbol{\beta}$. Note that this $\mathbf{\tilde{Y}}$ is different from $\mathbf{Y}$ described above. $\mathbf{Y}$ contains $N$ runs and is required to estimate coefficients using the model selection algorithms.  $\mathbf{\tilde{Y}}$ is generated to evaluate the selected models from each algorithm. We then calculate a new ``prediction'' response using the equation of the fitted model by $\mathbf{\hat{Y}}_i = \mathbf{X_{SFD}'}\boldsymbol{\hat{\beta}}$.} For both $\mathbf{\tilde{Y}}$ and $\mathbf{\hat{Y}}$ we have the benefit of having 1,000 response values calculated using the entire design region as opposed to using only the 17 response values in a DSD with $K = 4$. Figure \ref{fig:DSD_SFD_Cube} illustrates the difference in the number of comparison points between the original design and the 1,000 points generated from the SFD.

We calculate the RMSPE scores for all 1,000 $\mathbf{\hat{Y}}_i$'s for each model algorithm and noise level and present the results in Section \ref{sec:sim_results}. Due to extreme outlier RMSPE scores (in simulations mostly featuring the fit definitive screening procedure), we display the log(RMSPE). The RMSPE results can be found in the supplementary materials.

\subsection{Simulation Results}\label{sec:sim_results}

\begin{figure}[H]
\centering
\hspace*{-0.6cm}
\includegraphics[width=1\textwidth, angle=0]{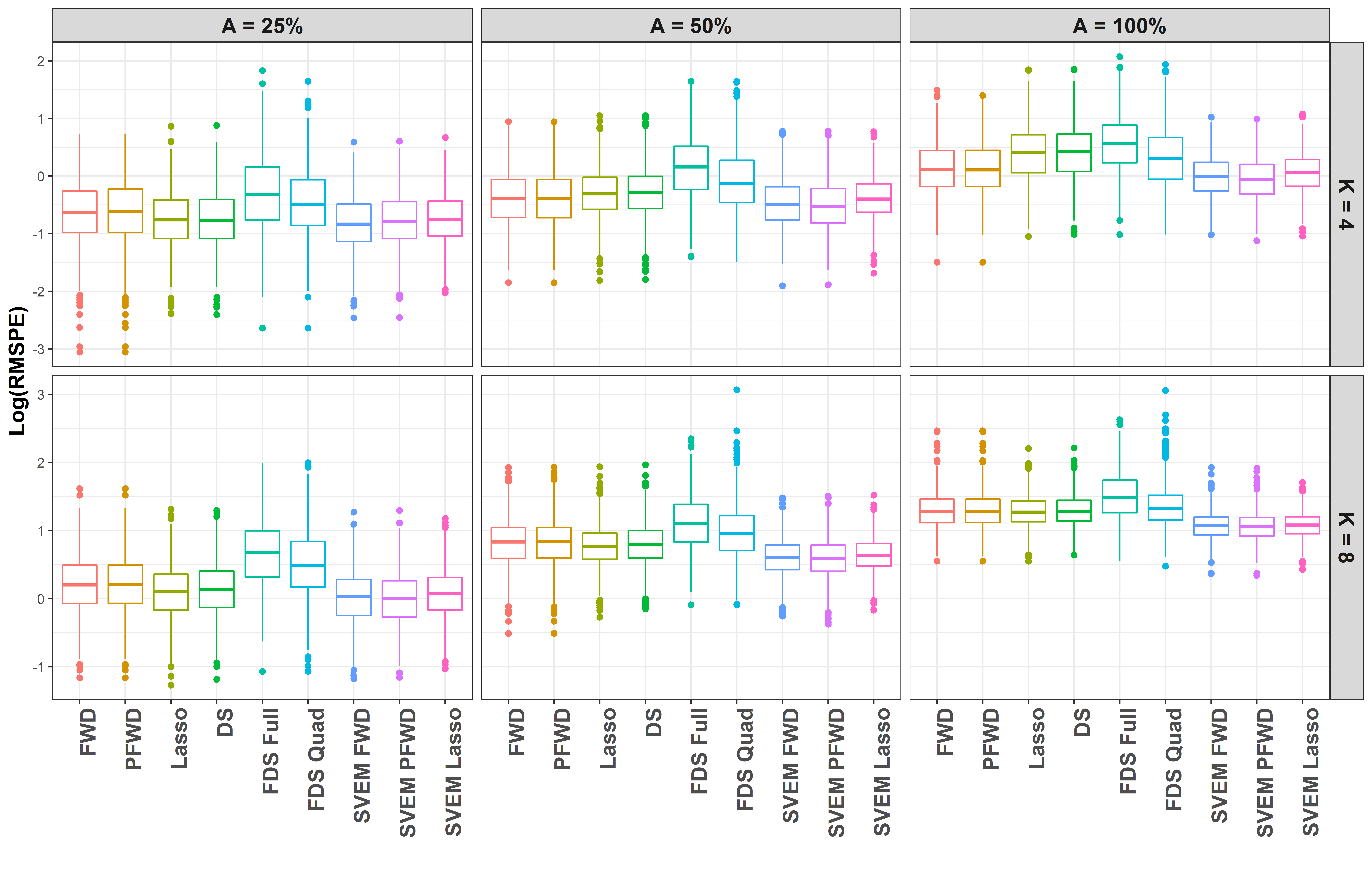}
\caption{DSD log(RMSPE) simulation results when $\sigma = 1$}\label{fig:DSD_boxplots_1}
\end{figure}

\begin{figure}[H]
\centering
\hspace*{-0.6cm}
\includegraphics[width=1\textwidth, angle=0]{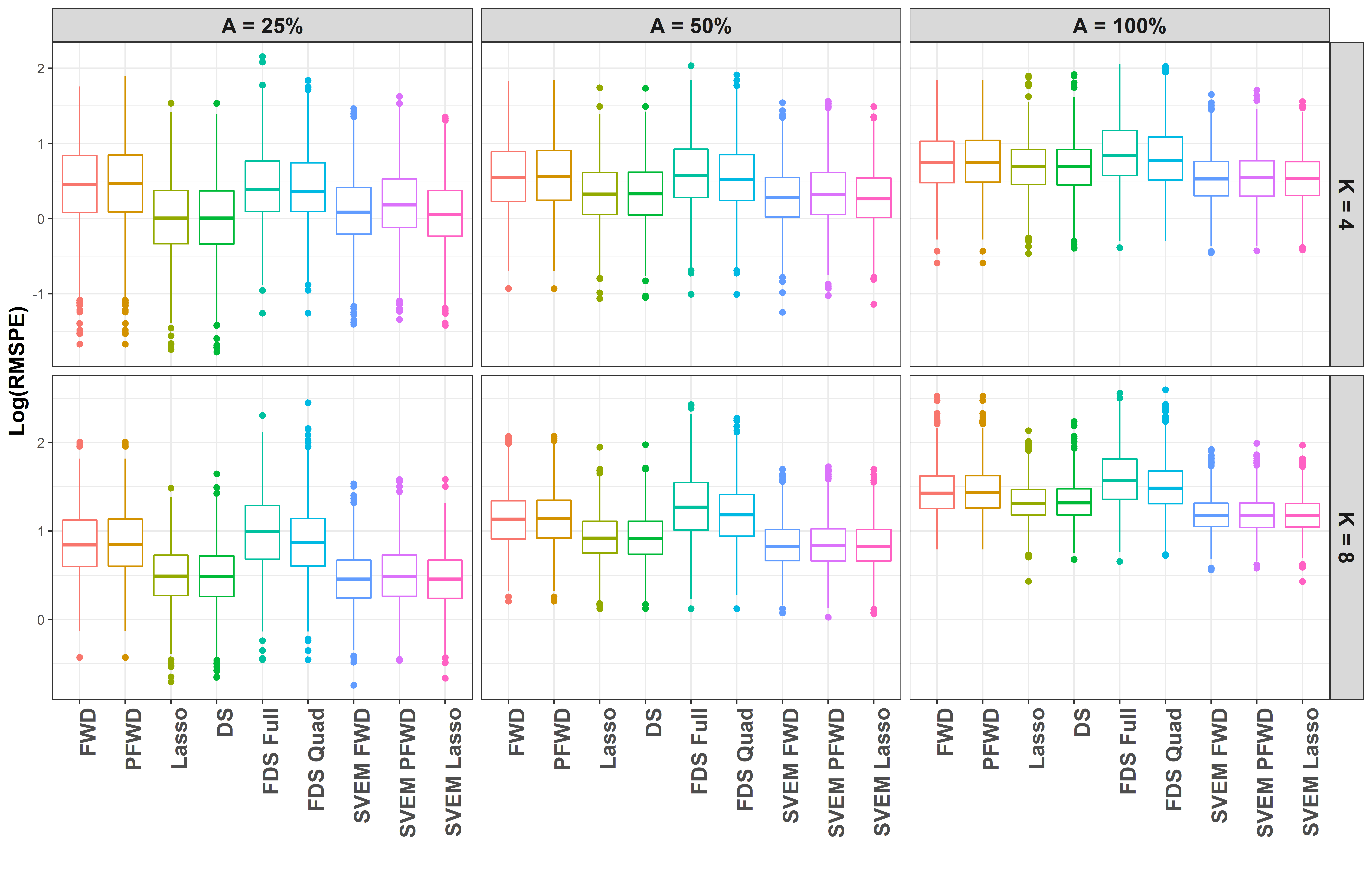}
\caption{DSD log(RMSPE) simulation results when $\sigma = 3$}\label{fig:DSD_boxplots_3}
\end{figure}

For the DSDs in Figures \ref{fig:DSD_boxplots_1} and \ref{fig:DSD_boxplots_3}, we see a clear pattern emerging: SVEM model algorithms perform better relative to their non-SVEM counterparts when there is less sparsity imposed in the true model. Conversely when A = 25\%, in both the $\sigma = 1$ and $\sigma = 3$ case we find a fair amount of overlap between all the model selection methods; however in every scenario, the fit definitive screening has higher RMSPE values. On balance we see the SVEM methods performing consistently between the three versions with the forward selection implementation performing the best overall. 

\begin{figure}[H]
\centering
\hspace*{-0.6cm}
\includegraphics[width=1\textwidth, angle=0]{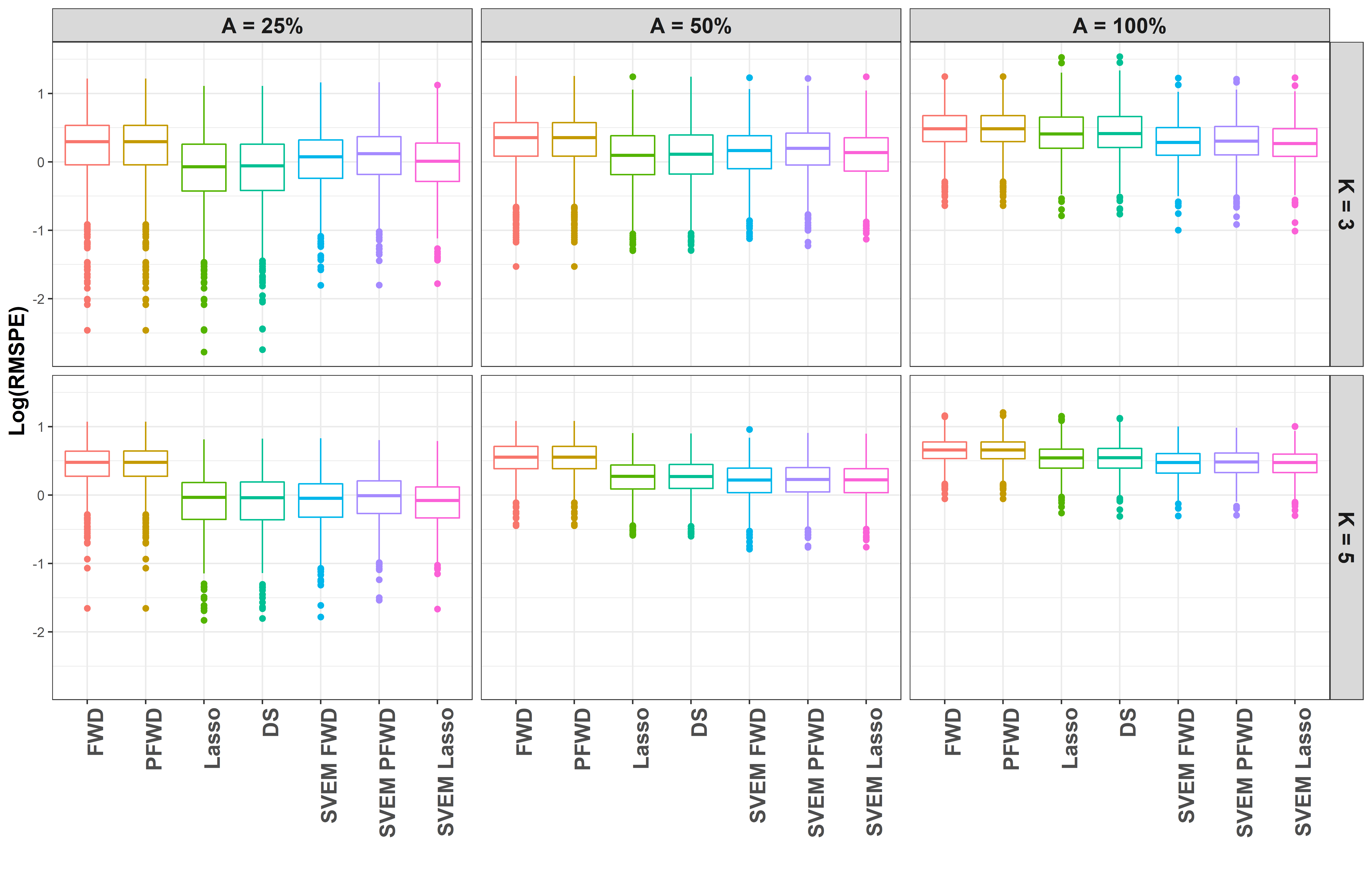}
\caption{BBD log(RMSPE) simulation results when $\sigma = 1$}\label{fig:BBD_boxplots_1}
\end{figure}

\begin{figure}[H]
\centering
\hspace*{-0.6cm}
\includegraphics[width=1\textwidth, angle=0]{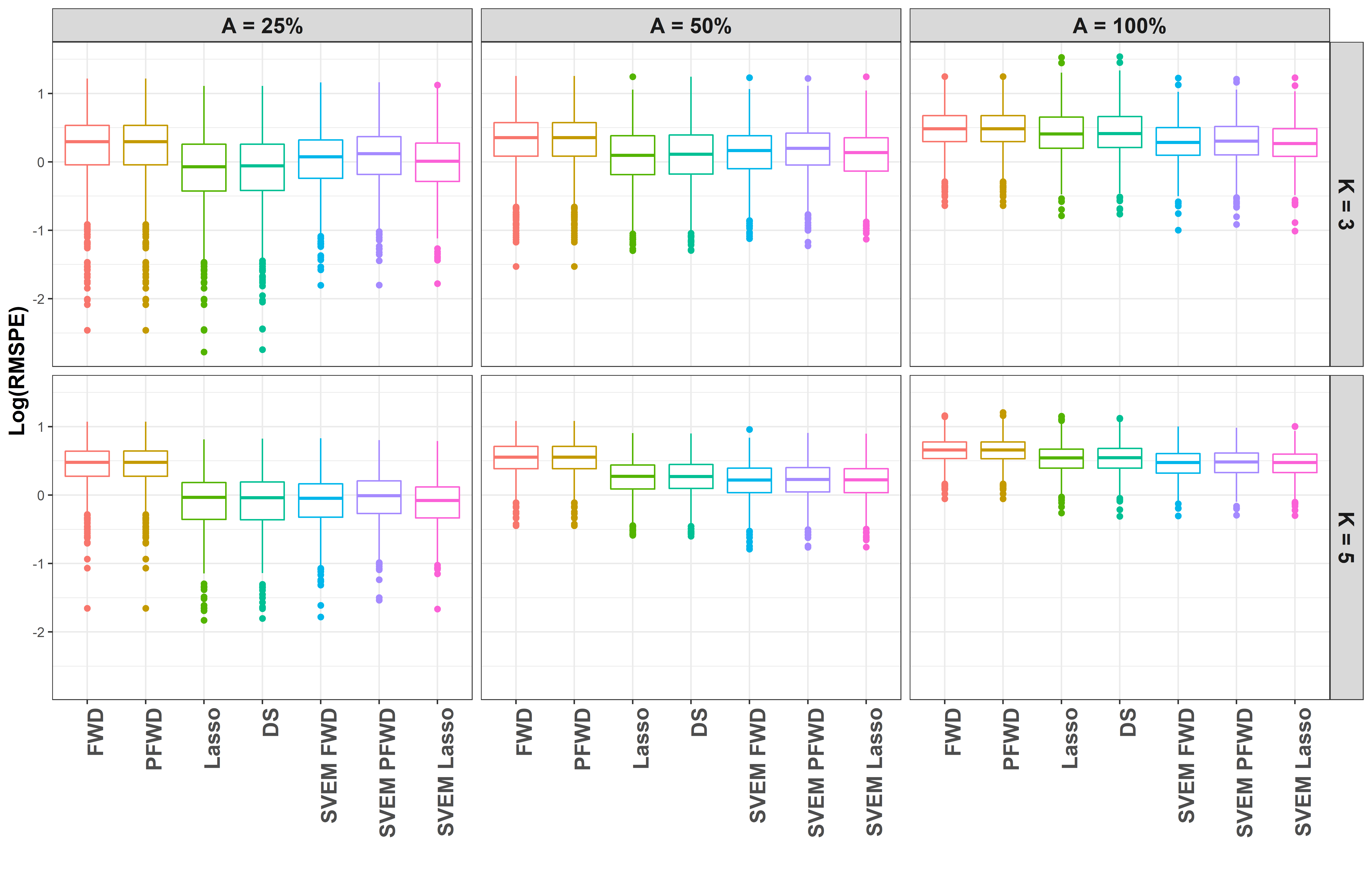}
\caption{BBD log(RMSPE) simulation results when $\sigma = 3$}\label{fig:BBD_boxplots_3}
\end{figure}

The the BBD simulation results, presented in Figures \ref{fig:BBD_boxplots_1} and \ref{fig:BBD_boxplots_3} show a similar trend to the DSD results but the contrast in performance of the SVEM model algorithms are not as stark. We remind the reader that in the BBD simulations, each design has $N>P$ i.e. none of the model matrices were super-saturated (like the DSD with K = 8) or saturated (like the DSD with K = 4). In Figure \ref{fig:BBD_boxplots_1} where $\sigma = 1$, the SVEM results are only slightly better than the Lasso and DS when A = 100\% for both K = 3 and K = 5. When A = 25\% and 50\%, we see no appreciable difference between SVEM and the Lasso and DS.

In Figure \ref{fig:BBD_boxplots_3} where $\sigma = 3$, \textcolor{black}{similar to} Figure \ref{fig:BBD_boxplots_1}, the SVEM models have slightly improved median log(RMSPE) scores compared to the Lasso and the DS. The forward selection and pruned forward selection model algorithms have the highest log(RMSPE) scores in all cases. In all other sparsity scenarios (Figure \ref{fig:BBD_boxplots_3}) for both K = 3 and K = 5, we only notice one case (A = 25\% and K = 3) where the Lasso and DS have slightly lower median log(RMSPE) scores than all other model algorithms, most notably the SVEM model algorithms, particularly SVEM with forward selection and SVEM with Lasso, which are in a close second position.

\textcolor{black}{From these simulations we can conclude that} SVEM outperforms its non SVEM counterparts when the true model has very low sparsity. The results become even more apparent 
when smaller designs are used, as seen in Figures \ref{fig:DSD_boxplots_1} and \ref{fig:DSD_boxplots_3}. We further see this trend exacerbated when we increase the signal-to-noise ratio. 
\textcolor{black}{Even} when sparsity is high \textcolor{black}{in the true model}, we see no reason not to use SVEM since the results are on par with the best non-SVEM model, which is usually the Lasso and or the DS.   

\begin{figure}[H]
\centering
\hspace*{-0.6cm}
\includegraphics[width=1\textwidth, angle=0]{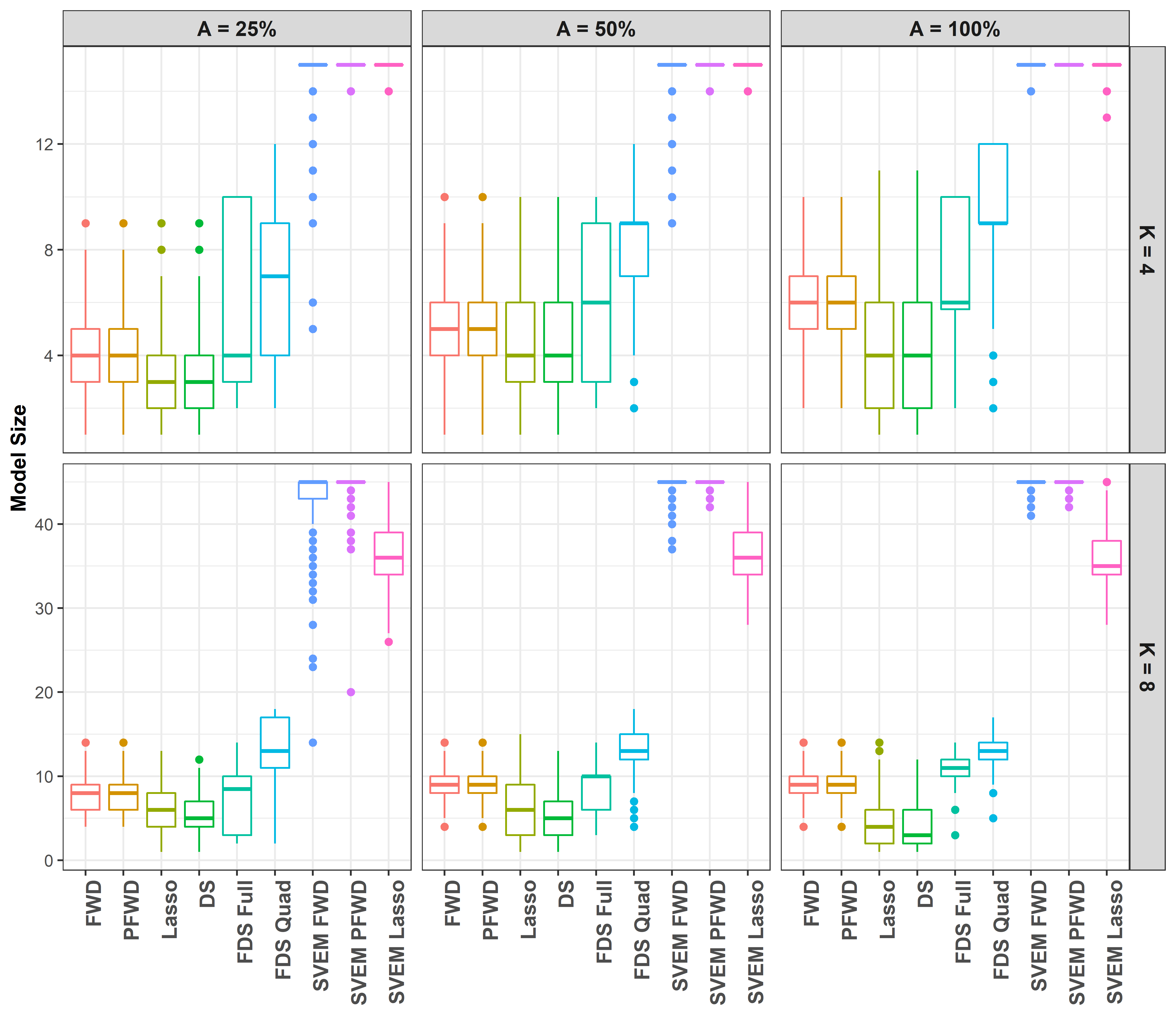}
\caption{DSD simulation model size results when $\sigma = 1$}\label{fig:DSD_boxplots_modsize_1}
\end{figure}

To gain insight into the performance of SVEM, we show the average model size by analysis method for the DSDs when $\sigma=1$ (Figure \ref{fig:DSD_boxplots_modsize_1}). We find that most often SVEM will have $p\prime = P$. 
Even in the case of the simulations with a DSD with $K = 8$ (Figure \ref{fig:DSD_boxplots_modsize_1}), we find this to be true even though $P = 45$ and $N = 21$. SVEM therefore produces a final model where the total number of selected effects exceeds the degrees of freedom available, 
increasing fitted model complexity but subsequently improving predictive ability. In the case of the BBD simulations and when $\sigma=3$, we see a similar pattern in model size (see supplementary materials for additional figures). 

\begin{figure}[H]
\centering
\hspace*{-0.6cm}
\includegraphics[width=1\textwidth, angle=0]{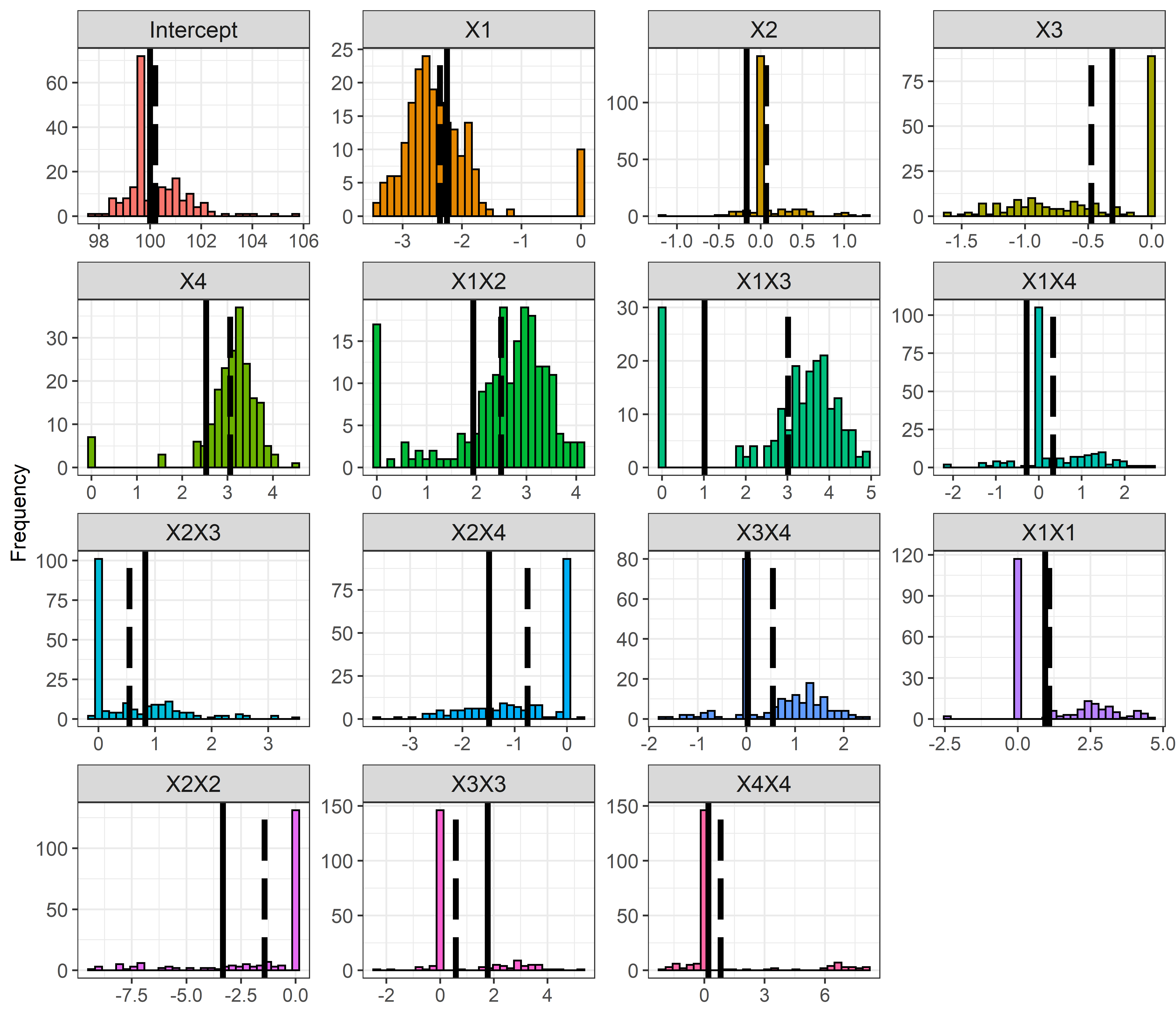}
\caption{Distributions of coefficients from a single DSD (K = 4, A = 100\%, $\sigma = 3$) simulation using FWD with SVEM and nBoot = 200. The solid vertical line represents the true model coefficient value and the dashed vertical line represents the final SVEM coefficient value.}\label{fig:DSD_4_Hists}
\end{figure}

Even though the final model size for SVEM models appears to be large, it is important to note that some of the individual coefficients might be quite small. In Figure \ref{fig:DSD_4_Hists} we display the distributions of the coefficients for each predictor from a single simulation run in the DSD $K = 4$, A = ALL, and $\sigma = 3$ scenario. In this example, we have $P\approx N$, and a \textcolor{black}{model matrix in which every column} has a non-zero coefficient value. All 14 predictors are selected at least once and the coefficient has a non-zero value in every $nBoot$ run. Some predictors have been selected far more than others ($X_1, X_4, X_1X_2, X_1X_3$), while the majority fall into the category of moderate to slight rate of selection. This is expected, based on the discussion of \cite{burn_and} 
stating that predictors do not have a binary contribution to a model but, rather, all predictors have some effect size, albeit some will have an effect size near zero ($X_4X_4$).




We ran an additional simulation to assess the optimal number for the $nBoot$ hyperparameter. We used a DSD with $K = 8$, A = ALL, and $\sigma = 3$. We utilized the same protocol as presented in Section \ref{sec:sim_protocol}. 
We used only SVEM with forward selection since it had the best performance on average (see Section \ref{sec:sim_results}). We iterated over eight values of $nBoot$: 1, 50, 100, 200, 500, 1000, 2000, and 3000. We observed no significant improvement by increasing $nBoot$ above 200.  We recommend $nBoot$ of 200 with an upper limit of 500. A user may wish to employ more $nBoot$ runs, but there is unlikely to be an appreciable improvement in SVEM's performance (see supplementary materials). 



\section{A Case Study: Plasmid Manufacturing}\label{sec:casestudy}

\subsection{Overview}

Plasmids (or pDNA) are non-chromosomal, circular-shaped DNA found in the cells of various types of bacteria (e.g., E.coli).  Plasmids are a key component in many new, biologic-based cell and gene therapeutics, resulting in an increased demand for high-quality plasmids. However, plasmid manufacturing is challenging and demand for plasmids consistently exceeds the available supply \citep{Xenopoulos_pDNA}. 

\subsection{Experimental Design and Models}

In this case study a \textcolor{black}{$K=5$, $N=15$}, DSD was used to study the fermentation step of a plasmid manufacturing process. The goal of the experiment is to build a predictive model that is subsequently used to characterize and optimize a fermentation step of the plasmid manufacturing process. In addition, \textcolor{black}{a separate and independent experiment using a} \textcolor{black}{$K=5$, $N=31$} central composite design (CCD) was performed. The experiment based on the CCD was performed independently of the DSD experiment in order to provide an independent assessment of the DSD predictive model performance. Tables \ref{tab:pDNA_CCD} and \ref{tab:pDNA_DSD} in the Appendix display the data for the two experiments. Note that generic settings of the five factors are presented because the actual settings are proprietary. However, the values of the response, pDNA Titer mg/L (Yield), are the actual \textcolor{black}{experimental} values observed.

\textcolor{black}{In this case study}, we use various model selection methods, including SVEM, to build a predictive model for pDNA Titer based on data \textcolor{black}{collected} from the DSD \textcolor{black}{experiment}, and then apply \textcolor{black}{the subsequent} predictive model to the \textcolor{black}{experimental} data \textcolor{black}{collected} from the CCD study in order to obtain an independent assessment of prediction performance on a true test set. 



\subsection{Results and Discussion}

\textcolor{black}{To evaluate the performance of SVEM and other model selection procedures we compare predictions made from the model constructed using the DSD experiment to the results from the CCD experiment which are considered the “true” values. We compare SVEM paired with Forward Selection and Lasso to classical “one-shot” approaches using both BIC and AICc with forward selection, and Lasso. Lasso with AICc was dropped since Lasso resulted in an intercept-only model, which neither predicts well nor serves any value for the study. In addition, the p-value based fit definitive method of \cite{fitdef} was applied to the DSD data for model selection. In all cases, only FQ models were considered. }

\begin{table}[h]
\begin{tabular}{llcccc}
\hline
\textbf{Criterion} & \textbf{\begin{tabular}[c]{@{}l@{}}Model \\    Selection\end{tabular}} & \multicolumn{1}{l}{\textbf{\begin{tabular}[c]{@{}l@{}}Full Model\\    (p = 20)\end{tabular}}} & \textbf{\begin{tabular}[c]{@{}c@{}}RMSPE\\ DSD\end{tabular}} & \textbf{\begin{tabular}[c]{@{}c@{}}RMSPE\\ CCD\end{tabular}} & \textbf{$\mathbf{R^2}$ CCD} \\ \hline
SVEM              & \begin{tabular}[c]{@{}l@{}}Forward\\ Selection\end{tabular}            & FQ                                                                                            & 15.7                                                        & 53.5                                                        & 0.55                              \\
SVEM              & Lasso                                                                  & FQ                                                                                            & 14.8                                                        & 57.8                                                        & 0.45                              \\
BIC                & Lasso                                                                  & FQ                                                                                            & 16.0                                                        & 61.4                                                        & 0.44                              \\
AICc                & \begin{tabular}[c]{@{}l@{}}Forward\\ Selection\end{tabular}          & FQ                                                                                            & 29.4                                                        & 71.5                                                        & 0.50                              \\
P-Values           & Fit Definitive                                                         & FQ                                                                                            & 40.0                                                        & 64.8                                                        & 0.35                              \\ 
BIC                & \begin{tabular}[c]{@{}l@{}}Forward\\ Selection\end{tabular}          & FQ                                                                                              & 16.5                                                        & 76.4                                                        & 0.46                              \\
\hline
\end{tabular}
\caption{\label{tab:pDNA_Results}Prediction performance on the CCD for all analysis methods.  Note that here $nboot$ = 1,000.}
\end{table}

\begin{figure}[H]
\centering
\hspace*{-0.6cm}
\includegraphics[width=0.8\textwidth, angle=0]{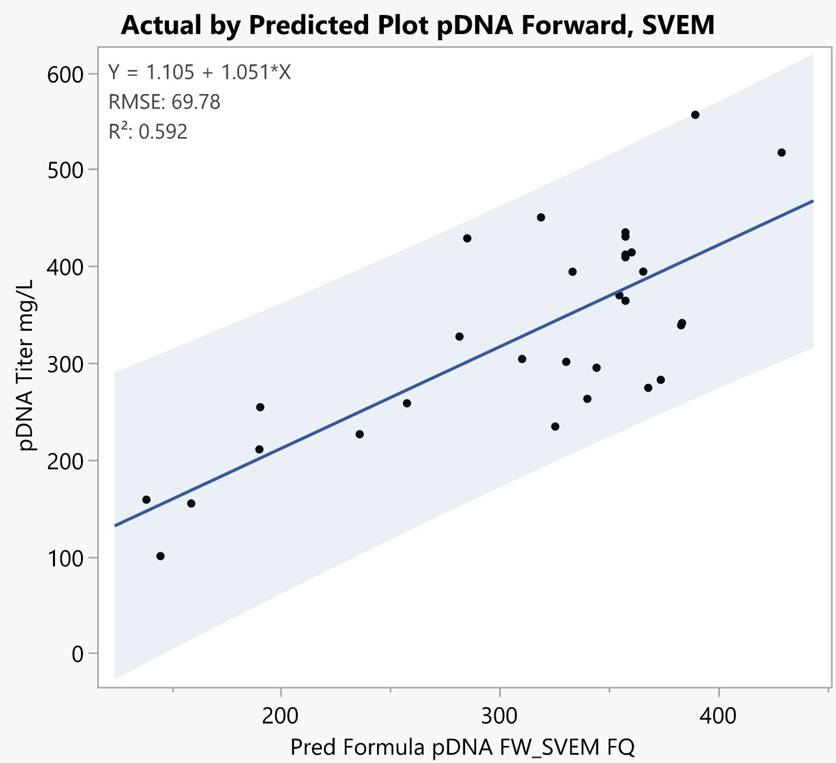}
\caption{Predicted by actual plot. The CCD pDNA values plotted against the predicted pDNA values from the four best performing model algorithms.}
\label{fig:pDNA_CCD_RASE_pred_plots}
\end{figure}

Table \ref{tab:pDNA_Results} displays \textcolor{black}{RMSPE and $R^2$ on calculated on the response from CCD experiment}. The actual by predicted plot for the best performing model algorithm (SVEM with forward selection) is displayed in Figure \ref{fig:pDNA_CCD_RASE_pred_plots}. The smallest test set RMPSE values were achieved with the SVEM-based methods. From Table \ref{tab:pDNA_Results} we find that SVEM with forward selection achieved the smallest RMPSE = 53.5, while second-best results were for SVEM with Lasso. All the non-SVEM methods had test set RMPSE values $>$ 60.0. The worst performers in terms of test set RMPSE were forward selection with AICc, fit definitive, and forward selection with BIC. As mentioned, the Lasso with AICc criterion resulted in an intercept-only model despite searching for an optimal $\lambda$ (Lasso shrinkage parameter) over a 1,500 point grid on a square root scale and a log scale.
                

Often, a study’s goal is to characterize the entire design region with the intent on scaling up a process. In these cases, the larger SVEM models are preferred for two reasons: they have the capacity to deal with the complexity over the entire experimental region \textcolor{black}{(Figure \ref{fig:pDNA_CCD_RASE_pred_plots})}, and the ability to accommodate the increased process complexity often encountered in scale-up activities.

\section{Concluding Remarks} \label{sec:conclusion}

In this paper we have proposed \textcolor{black}{SVEM, a general algorithm} that allows for ensemble modeling in small $N$ situations, typical of designed experiments. We demonstrated via simulations and a case study that the use of SVEM leads to lower average prediction error on independent test sets when compared to its non-SVEM counterparts. \textcolor{black}{This is most apparent} in cases of low sparsity and when the signal-to-noise ratio is small and when $N$ is small. \textcolor{black}{SVEM makes minimal model building assumptions that are typically associated with statistical models: i.i.d normally distributed errors and $p\prime < N-1$ are not required by SVEM.} 
SVEM is very general in its approach and could be applied to any designed experiment. 

 We suspect the reason for SVEM's improved performance over its non-SVEM counterparts relates to its ability to fit a supersaturated model via the ensembling procedure. It also overcomes the instability problem discussed by \cite{breiman_instability}. \textcolor{black}{SVEM shows us that we may gain greater accuracy by having more predictors in our models and not fewer.} \cite{Radford_Neal_Against_Parsimony} claims that sometimes simple models will outperform more complex models, but deliberately limiting the complexity of a model is not fruitful when the problem is evidently complex.

SVEM has opened the door to a wide array of related research topics. First is the issue of quantification of the uncertainty around the predictions. \textcolor{black}{Another important topic of future research is to study the design construction method that is best for use with SVEM. In principle, because SVEM has no distributional assumptions, it can be combined with non-normal response distributions, as well as more general ML model classes, such as neural networks, support vector machines, tree based algorithms etc. It is not yet known how well SVEM will work in the case of randomization restrictions where there are blocking or whole plot factors, or whether SVEM will need to be generalized to accommodate these structures.}

\textcolor{black}{SVEM's improved predictive accuracy does come at a cost. If the underlying model algorithm is computationally demanding (like best subsets or support vector machines) then the computation time for SVEM and, say, 200 $nBoot$ runs will be rather demanding on time and computational power. }

\section{Acknowledgements}\label{sec:acknowledgements}

We would like to thank John Sall, Clay Barker, Marie Gaudard and the two anonymous reviewers for their excellent guidance in the presentation of this work.

\bibliographystyle{apalike} 
\bibliography{main.bib}

\section{Appendix}





\begin{table}[!htbp]
\begin{tabular}{cccccc}
\hline
\textbf{pH} & \textbf{\%DO} & \textbf{\begin{tabular}[c]{@{}l@{}}Induction Temperature\\ \ \ \ \ \ \ \ \ \          (Celsius)\end{tabular}} & \textbf{Feed Rate} & \textbf{\begin{tabular}[c]{@{}l@{}}Induction \\ \ \  OD600\end{tabular}} & \textbf{Titer mg/L} \\ \hline
1           & 1             & -1                                                                                           & 1                  & -1                                                                  & 581.36              \\
-1          & -1            & -1                                                                                           & 1                  & -1                                                                  & 519.80              \\
-1          & 1             & -1                                                                                           & -1                 & -1                                                                  & 115.40              \\
-1          & 1             & -1                                                                                           & 1                  & 1                                                                   & 407.22              \\
1           & 1             & -1                                                                                           & -1                 & 1                                                                   & 56.18               \\
1           & -1            & 1                                                                                            & 1                  & -1                                                                  & 260.82              \\
-1          & -1            & -1                                                                                           & -1                 & 1                                                                   & 94.95               \\
1           & -1            & -1                                                                                           & -1                 & -1                                                                  & 215.03              \\
-1          & 1             & 1                                                                                            & -1                 & 1                                                                   & 211.00              \\
0           & 0             & 0                                                                                            & 0                  & 0                                                                   & 321.00              \\
0           & 0             & 0                                                                                            & 0                  & 0                                                                   & 387.35              \\
-1          & 1             & 1                                                                                            & 1                  & -1                                                                  & 231.00              \\
1           & 1             & 1                                                                                            & 1                  & 1                                                                   & 351.00              \\
1           & -1            & -1                                                                                           & 1                  & 1                                                                   & 284.00              \\
-1          & -1            & 1                                                                                            & 1                  & 1                                                                   & 298.00              \\
-1          & -1            & 1                                                                                            & -1                 & -1                                                                  & 191.00              \\
1           & -1            & 1                                                                                            & -1                 & 1                                                                   & 183.02              \\
0           & 0             & 0                                                                                            & 0                  & 0                                                                   & 368.74              \\
1           & 1             & 1                                                                                            & -1                 & -1                                                                  & 111.46              \\
0           & 0             & 0                                                                                            & 0                  & 0                                                                   & 391.74              \\
0           & 0             & 0                                                                                            & 0                  & 0                                                                   & 366.01              \\
1.3         & 0             & 0                                                                                            & 0                  & 0                                                                   & 257.88              \\
-1.3        & 0             & 0                                                                                            & 0                  & 0                                                                   & 295.68              \\
0           & 1.3           & 0                                                                                            & 0                  & 0                                                                   & 385.54              \\
0           & -1.3          & 0                                                                                            & 0                  & 0                                                                   & 371.02              \\
0           & 0             & 1.3                                                                                          & 0                  & 0                                                                   & 326.70              \\
0           & 0             & -1.3                                                                                         & 0                  & 0                                                                   & 251.76              \\
0           & 0             & 0                                                                                            & 1.3                & 0                                                                   & 351.11              \\
0           & 0             & 0                                                                                            & -1.3               & 0                                                                   & 167.39              \\
0           & 0             & 0                                                                                            & 0                  & 1.3                                                                 & 219.64              \\
0           & 0             & 0                                                                                            & 0                  & -1.3                                                                & 239.29              \\ \hline
\end{tabular}
\caption{CCD for the pDNA yield study. pH is a negative log concentration and the induction temperature is in celsius.}
\label{tab:pDNA_CCD}
\end{table}

\begin{table}[htb]
\begin{tabular}{cccccc}
\hline
\textbf{pH} & \textbf{\%DO} & \textbf{\begin{tabular}[c]{@{}l@{}}Induction Temperature\\ \ \ \ \ \ \ \ \ \          (Celsius)\end{tabular}} & \textbf{Feed Rate} & \textbf{\begin{tabular}[c]{@{}l@{}}Induction \\ \ \  OD600\end{tabular}} & \textbf{Titer mg/L} \\ \hline
0  & 1    & 1                       & -1                                                         & -1        & 156.20     \\
0  & -1   & -1                      & 1                                                          & 1         & 318.45     \\
1  & 0    & -1                      & -1                                                         & 1         & 398.00     \\
-1 & 0    & 1                       & 1                                                          & -1        & 285.60     \\
1  & -1   & 0                       & 1                                                          & -1        & 229.00     \\
-1 & 1    & 0                       & -1                                                         & 1         & 377.00     \\
1  & -1   & 1                       & 0                                                          & 1         & 290.00     \\
-1 & 1    & -1                      & 0                                                          & -1        & 123.00     \\
1  & 1    & 1                       & 1                                                          & 0         & 299.00     \\
-1 & -1   & -1                      & -1                                                         & 0         & 428.00     \\
0  & 0    & 0                       & 0                                                          & 0         & 327.80     \\
0  & 0    & 0                       & 0                                                          & 0         & 339.74     \\
0  & 0    & 0                       & 0                                                          & 0         & 387.35     \\
0  & 0    & 0                       & 0                                                          & 0         & 393.97     \\
0  & 0    & 0                       & 0                                                          & 0         & 348.08     \\ \hline
\end{tabular}
\caption{DSD for the pDNA yield study. pH is a negative log concentration and the induction temperature is in celsius.}
\label{tab:pDNA_DSD}
\end{table}

%
%
%

\end{document}